\documentclass[prl,aps,twocolumn,dcolumn,superscriptaddress,amsmath,amssymb]{revtex4}

\usepackage{graphicx}
\usepackage{graphicx}
\begin{document}
\title{Fern\'andez and Alonso reply:}

\maketitle
Our conclusions \cite{Letter}, about the value of $p$ in the time 
evolution $m\propto t^p$ of the magnetization, are questioned
in Ref.  \cite{TS}. In order to dispel the suspicion that our Monte 
Carlo (MC) results
may be size dependent, we show in Fig. 1 results for various system 
sizes, both in SC and FCC lattices, that (1) exhibit that $p$ does
depend on lattice structure, as reported in Ref. \cite{Letter}, and 
(2) show no trace of any size dependence. We find size effects
only in smaller systems.

The variation of $p$ with lattice structure should not be perplexing. 
It is predicted by our theory \cite{PRB}, of which the
main result, i.e., Eq. (11) of Ref. \cite{Letter}, is known. Its 
predictions, given in Ref. \cite{Letter}, are also shown in Fig. 1.
Furthermore, for vanishing tunneling window widths, Eq. (11) of Ref. 
\cite{Letter} gives \cite{PRB}
\begin{equation}
\frac{\sin (p\pi )}{p}=\sqrt{2\pi}\frac{\sigma}{h_0}
\label{1}
\end{equation}
for {\it all} times well after $\Gamma^{-1}$, where $\sigma$ is 
approximately the dipole field rms value \cite{PRB},
$h_0$ is $8\pi^2/3^{5/2}$ times a nearest neighbor dipole field, and
$\Gamma^{-1}$, as well as all other notation in this Reply, are as 
defined in Ref.
\cite{Letter}. Values for
$\sigma$ are given in Ref. \cite{PRB} for various lattices, but, for 
completeness' sake, we give here $\sigma/h_0=0.756$, $0.398$,
$0.417$, and
$0.66$ for SC, BCC, FCC, and Fe$_8$ lattices, respectively. Values 
for $p$ that follow from Eq. (1) agree rather well with the
MC results we have reported and with experiments on Fe$_8$ \cite{ww}.

\begin{figure}
\includegraphics*[width=7.5cm]{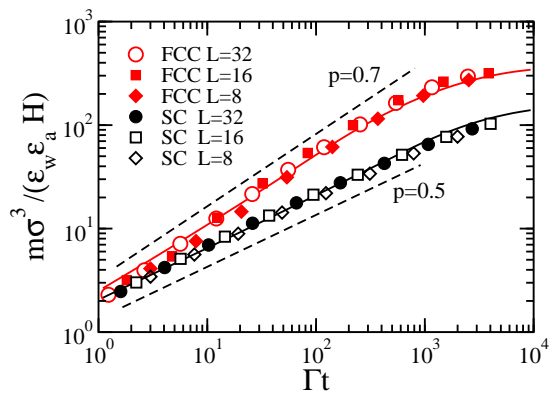}
\caption{$m$, scaled with $\varepsilon_w\varepsilon_aH/\sigma^3$, 
versus $\Gamma t$ for SC
and FCC lattices with
$N=L\times L\times L$ and
$4\times L\times L\times L/2$ spins, respectively. All points stand 
for averages over at least $10^7/N$ MC runs, with
$\varepsilon_w=0.05$, $H=1$, and $\varepsilon_a=0.43$ and $0.50$, for 
SC and FCC lattices, respectively. Full lines stand for
theoretical results.}
\label{nvst}
\end{figure}

So why are the numerical results of the Ref. \cite{TS} so different 
from ours? Ours \cite{Letter} apply to (1)
{\it annealed} \cite{annealed} systems, to which
a magnetic field $H$ is applied at $k_BT\lesssim 0.1U/S$, if (2) 
$H\ll \sigma$ [such as
$H\lesssim 4$ mT and
$\sigma\approx 30$ mT (i.e., $g\mu_BS\mu_0H\lesssim 0.05$ K and 
$g\mu_BS\mu_0\sigma\approx 0.4$ K in obvious notation) in Ref.
\cite{ww}]. Neither of these two conditions is met in Ref \cite{TS}, 
where initial spin configurations are random and $H>\sigma$.

Annealing is essential in the magnetization process studied in Ref. 
\cite{Letter}. The very
nature of the process depends on it. In annealed systems, spin--up 
and spin--down populations that are
able to tunnel, i.e., on sites where dipolar fields approximately 
cancel $H$, are unequal, and this drives
the magnetization process. In unannealed systems both populations 
are, on the average, initially equal, and much slower
thermal equilibration processes then drive the magnetization 
evolution \cite{otro}.
Furthermore, annealing is hard to avoid in experiments such as in 
Ref. \cite{ww}. An Fe$_8$ crystal held for as little as 1 second
within the range $20\gtrsim T\gtrsim 2$ K before quenching to much 
lower temperatures,
will qualify as annealed \cite{annealed}, and its magnetization will 
be at least two orders of magnitude larger, for up to minute
(roughly $4\Gamma^{-1}$) after $H\ll\sigma$ is applied, than it would 
have been had the initial spin configuration been somehow
completely randomized initially.

Furthermore, the results we have obtained for the magnetization of 
annealed systems also apply to the relaxation of the magnetization
in zero field, after cooling in a {\it weak} field \cite{PRB}. Then, 
$m(0)-m(t)\propto t^p$ while
$1\lesssim \Gamma t$ and $m(t)\gtrsim 0$, and $p$ is also given by Eq. (1).

Incidently, in
unannealed systems (as in Ref. [1]), $p$ is also nonuniversal, 
varying with $H$, increasing monotonically as $H$
decreases, up to a value slightly larger than 1 for $H\ll\sigma$.

Financed through grant BFM2003-03919-C02-01/FISI of the MCYT of Spain.

\noindent
J. F. Fern\'andez$^{1}$, and J. J. Alonso$^{1,2}$ \\
$^{1}$ ICMA, CSIC and UZ, 50009-Zaragoza, Spain\\
$^{2}$F\'{\i}sica Aplicada I, UMA, 29071-M\'alaga, Spain \\


\begin{thebibliography}{99}
\bibitem{annealed}By ``annealing'' we mean a significant amount of 
thermal equilibration at $T\gtrsim 0.1U/k_B$ and well above the
magnetic ordering temperature.
\end{thebibliography}
\end{document}